# AffectiveSpotlight: Facilitating the Communication of Affective Responses from Audience Members during Online Presentations

Prasanth Murali
Northeastern University
Boston, MA, USA
murali.pr@northeastern.edu

Javier Hernandez
Microsoft Research
Cambridge, MA, USA
javierh@microsoft.com

Daniel McDuff
Microsoft Research
Cambridge, MA, USA
damcduff@microsoft.com

Kael Rowan
Microsoft Research
Redmond, WA, USA
kaelr@microsoft.com

Jina Suh
Microsoft Research
Redmond, WA, USA
jinsuh@microsoft.com

Mary Czerwinski
Microsoft Research
Redmond, WA, USA
marycz@microsoft.com

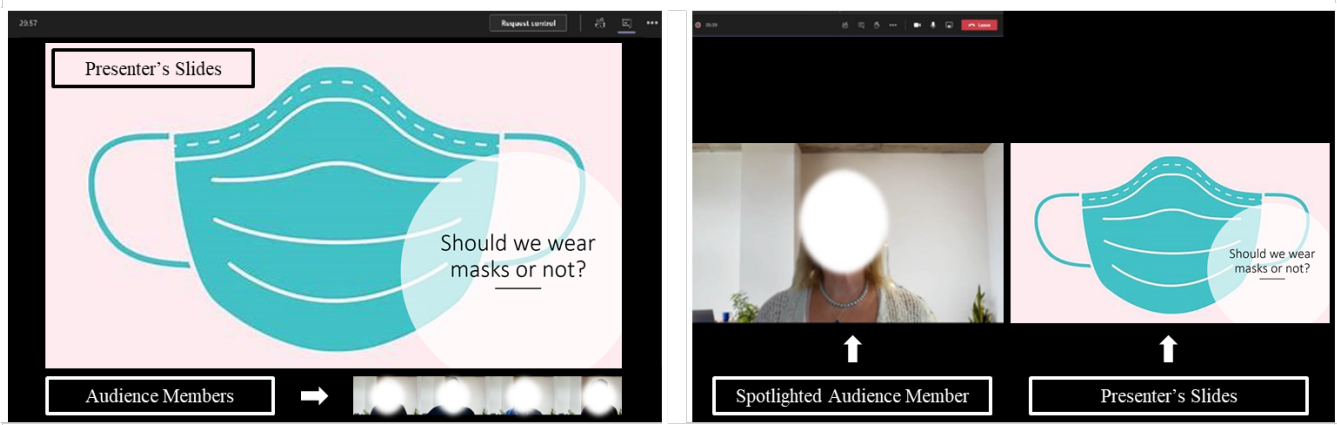

Figure 1: Existing videoconferencing platforms such as Microsoft Teams provide limited audience feedback when giving presentations (left). We develop AffectiveSpotlight which analyzes and spotlights audience members in real-time to support presenters (right). Text boxes indicate the main components of the platforms.

## ABSTRACT
The ability to monitor audience reactions is critical when delivering presentations. However, current videoconferencing platforms offer limited solutions to support this. This work leverages recent advances in affect sensing to capture and facilitate communication of relevant audience signals. Using an exploratory survey (N=175), we assessed the most relevant audience responses such as confusion, engagement, and head-nods. We then implemented AffectiveSpotlight, a Microsoft Teams bot that analyzes facial responses and head gestures of audience members and dynamically spotlights the most expressive ones. In a within-subjects study with 14 groups (N=117), we observed that the system made presenters significantly more aware of their audience, speak for a longer period of time, and self-assess the quality of their talk more similarly to the audience members, compared to two control conditions (randomly-selected spotlight and default platform UI). We provide design recommendations for future affective interfaces for online presentations based on feedback from the study.

## CCS CONCEPTS
• **Human-centered computing** → **Human computer interaction**; *Interactive systems and tools*; **User interface programming**;

## KEYWORDS
Affective Computing, Public Speaking, Intelligent User Interfaces, Videoconferencing

### ACM Reference Format:
Prasanth Murali, Javier Hernandez, Daniel McDuff, Kael Rowan, Jina Suh, and Mary Czerwinski. 2021. AffectiveSpotlight: Facilitating the Communication of Affective Responses from Audience Members during Online Presentations. In *CHI Conference on Human Factors in Computing Systems (CHI '21), May 8–13, 2021, Yokohama, Japan.* ACM, New York, NY, USA, 13 pages. https://doi.org/10.1145/3411764.3445235

## 1 INTRODUCTION
Giving presentations is a necessary part of many jobs such as teaching, management, and sales [1–4], and requires presenters to continuously gauge audience responses to intervene and ensure that the message is being communicated effectively. For instance, a presenter might inject humor or provide further clarifications upon observing a bored or confused audience, respectively. In contrast to in-person presentations, however, online presentations provide limited to no audience feedback, making it very difficult to establish rapport and effectively adapt the content of the presentation. This problem is partly due to the constraints of existing video-conferencing platforms, which often prioritize showing the presenter slides and/or show only a limited number of participants. Participants in a video call also turn off their cameras, which omit non-verbal cues that are useful to the presenter. In addition, remote presenters often have limited space on their computer displays, severely constraining the potential communication bandwidth. Due to the recent increase of remote work associated with the pandemic demands, this problem is more prevalent now than ever before.

Public speaking is often regarded as one of the most stressful daily activities [5] and is heavily influenced by audience responses to the presenter. In fact, studies that seek to reliably induce acute stress on people often involve giving a presentation in front of a neutral-looking audience (a.k.a., Trier social stress test [6]). While research on audience responses in online settings is still nascent, there is prior work considering the impact of in-person audience responses [7], especially in the context of alleviating public speaking anxiety. For instance, MacIntyre and Thivierge [8] showed that low perceived audience interest, responsiveness, and evaluation of the talk can contribute to public speaking anxiety. Other studies [9–11] have identified that high audience responsiveness, in terms of head nods and smiling, induced less anxiety and promoted more communication. These findings are consistent with Motley's work [12], that proposed a continuum for presenter's orientations based on audience reactions, ranging from low to high audience responsiveness, interest, and evaluative stance towards the presenter. The limitations of current video-conferencing applications such as the lack of an intuitive reception of audience feedback, could potentially make online presentations fall at the lower end of Motley's continuum, leading to a negative presenter experience. Thus, in this work, our research goal is to address the problem of limited access to audience responsiveness during online presentations to improve presenter's awareness of the audience by spotlighting reactive audience members to the presenter as they speak.

The spotlight metaphor was inspired by current approaches in theater and cinematography, in which camera recorders often highlight audience responses to capture the most relevant moments [13, 14], draw attention to elements on screen, and potentially evoke emotional responses [15]. Beyond cinematic experiences, the spotlight technique has also been used in computer applications to direct and maintain users' attention while simultaneously making surrounding context still visible [16–19]. Moreover, Khan et al. [20] showed that spotlighting can also help manage users' attention between their primary activity and peripheral information, even on small monitors. Since oral presentations constitute a similar paradigm requiring focused attention across the presentation slides, speaking notes, and the audience feedback, we embraced the spotlight analogy as a design solution to provide audience feedback to presenters in real-time.

Over the years, researchers have developed a wide variety of presenter-support systems to facilitate the sending and reception of both explicit and implicit feedback from the audience [21–25]. However, most of the work has focused on in-person presentations, in which presenters are often co-located with the audience and communication bandwidth is not limited. In contrast, we propose a presenter-support system that facilitates gathering implicit feedback from an online audience by leveraging recent advancements in computer vision-based affect sensing. In particular, we propose the AffectiveSpotlight system, which analyzes the facial responses and head gestures of audience members in real-time and dynamically spotlights the most expressive members for the presenter. We purposely decided to avoid labeling the inferred responses to empower presenters to make their own personal interpretations based on the context and their experience. To the best of our knowledge, our work is the first to explore the creation of an affect-driven spotlight that facilitates audience responses to online presenters to more closely resemble in-person presentation experiences.

This work is organized as follows. We first describe prior work in the context of facilitating audience feedback to presenters. We then describe an exploratory survey that helped identify what types of audience responses are most informative to presenters. Next, we use our findings from the survey to help inform the design and development of our AffectiveSpotlight system. We then describe a within-subject evaluation study that compared the proposed system with two other control presenter-support systems. We review the results of the study and provide design recommendations for future affective interfaces in the context of online presentations. Finally, we discuss our findings including the limitations and potential future directions.

## 2 RELATED WORK

Researchers have explored a wide variety of methods to enhance the sensing and communication of audience feedback for presenters. To help map the research in this space, Hassib et al. [26] identified four important dimensions: type of audience feedback (explicit vs. implicit), audience location (collocated vs. distributed), synchronicity of the feedback (synchronous vs. asynchronous), and sender/receiver cardinality (1 to 1, N to 1, and N to N). To help better position this work in the context of prior research, we review other work considering a person presenting to a large audience (i.e., 1:N cardinality).

Traditional methods of capturing audience feedback frequently rely on explicit modes such as use of self-reports and questionnaires, which are then aggregated and provided to the presenter in different ways. For instance, Rivera-Pelayo et al. developed the Live Interest Meter App [27], which is a mobile and desktop application that gathers responses from the audience on demand. In particular, the presenter introduces a question for the audience, such as comprehension level of the talk or speaking volume, and the application aggregates and displays a summary graph. In a separate work, Chamillard [28] explored the use of iClicker,[1] which enabled instructors to receive student responses during lectures. In this case, the researchers identified a strong relationship between student participation and learning. In the context of confusion, Glassman et al., [23] developed Mudslide, an anchored interaction tool that allowed the audience to indicate confusing points for different parts of online lectures. The instructors found that anchored feedback was more valuable and easier to interpret than feedback provided at the end of the lecture. In a separate work, Teevan et al. [25] designed a smartphone interface that enabled audience members to indicate thumbs up/down in real-time and visualize the feedback via a shared, projected visualization. The researchers identified that the system helped audience members pay close attention to the presentation, helped them stay connected to other audience members, and facilitated retrospective review of the session. Different forms of explicit feedback have also been adopted by some social media platforms, in which audience members can broadcast different reactions in the form of flying emojis (e.g., [29], [30]). Despite the benefits of explicit methods to capture audience feedback, it is commonly observed that they can also increase cognitive workload and distraction for both the presenter and audience members [25, 27]. In addition, there are scenarios where the audience members who are too attentive or distracted can forget to provide feedback.

---
[1] https://www.iclicker.com/

To help address these limitations, some studies have explored the use of implicit methods to capture audience responses, such as monitoring physiological or behavioral signals. In one of the earliest studies, Picard and Scheirer created Galvactivator [31], a hand-worn wearable that monitored the electrodermal responses of the audience members and increased the illumination of an LED when high physiological arousal was detected. The researchers observed that presenters found the information useful, especially for detecting both engaging and boring parts of the presentation. More recently, Hassib et al. developed EngageMeter [21], a head-mounted wearable that monitored electroencephalographic signals from the audience to estimate engagement and provided the feedback to the presenter in real-time. In this case, presenters found the information useful for knowing when to change the style of communication (e.g., tone of voice, injecting pauses). While physiological sensing is a promising technique, however, the cost and availability of custom sensors is a major obstacle, preventing the wide adoption of such technologies in real world scenarios. To address this problem, Sun et al. [32] developed a system that monitored facial expressions of students via webcam to estimate different cognitive states (e.g., anxiety, flow, boredom), and provided visualizations of the flow experience for the whole group to the instructor. The research identified the value of providing a real-time flow visualization, but also acknowledged that it still increased the cognitive load for the presenter.

Our work similarly considers the use of pervasive webcams to monitor the facial expressions and head gestures of the audience, but explores providing the information in the form of original video feeds, which may be more familiar and pose less cognitive demands to the presenter as opposed to aggregated data visualizations. This visualization approach is closely related to prior work (e.g., [33, 34]) that purposefully avoided labeling the sensed data to support flexibility in the interpretation, based on the context and personal experiences of the viewer. Our research goal is to identify relevant audience reactions that are most helpful to presenters and develop a system that spotlights audience members accordingly in the context of online presentations.

## 3 EXPLORATORY SURVEY

This section describes an exploratory survey to help us understand the current landscape of online meetings as well as relevant audience behaviors for presenters.

### 3.1 Methods

The survey included questions regarding job role and presentation habits, problems faced when presenting via videoconferencing systems, and preferred audience reactions and behaviors during online presentations. We asked the participants to report their frequency of giving presentations to gauge the level of presentation experience. We also asked them to compare their experiences for in-person and online presentations to understand challenges introduced by presentations being given online. To help identify the most relevant audience responses sought out by presenters, participants were then requested to rank a pre-selected list of behaviors and cognitive states. The different states were derived from feedback affordances in current videoconferencing systems (e.g., hand raise[2]) and prior research in computer vision-based affective computing (e.g., facial expression analysis [32, 33, 35]). Finally, to understand the acceptability for enabling cameras and to evaluate the feasibility of a video-based approach, we asked the participants to rate the likeliness of turning on the video camera as an audience member across different sizes of the audience.

An online survey was sent via e-mail to random members of a large technology company. We received a total of 175 responses from participants with job roles of Software Developers (27%), Engineers (25%), Program Managers (18%), Researcher (12%), Sales (10%) among others.

### 3.2 Results

We analyzed quantitative survey data using descriptive statistics to compare differences between the groups. For open-ended questions, the responses were coded by the lead author of the submission using thematic analysis techniques to draw insights [36, 37].

The majority of participants (83%) reported giving a presentation at least once a month, suggesting that our studied population had presentation experience. Participants also indicated that a large majority of the online presentations (41.25%) were given to an audience size between 5 and 10 members.

A large number of participants (83.11%) also reported missing relevant audience feedback when presenting online. In an optional open-ended question that compared available information for in-person vs. online presentations, participants indicated that current videoconferencing systems were limited in capturing three main types of audience feedback. Firstly, participants missed seeing the non-verbal social cues of audience members, which made it difficult to gauge presentation engagement, attention and focus. Secondly, participants reported missing the view of the audience, which enabled presenters to select specific audience members as well as "read the room" of in-person presentations. Thirdly, participants reported missing more active interactions with the audience (e.g., shared smiles) as it often felt like a one-way communication with no reflection of the audience energy level.

---

[2]https://support.microsoft.com/en-us/office/raise-your-hand-in-a-teams-meeting-bb2dd8e1-e6bd-43a6-85cf-30822667b372

When asked to rate the helpfulness of the audience responses most relevant to presenters, participants reported that they would like to see confusion and engagement states more than other cognitive states. Similarly, participants reported wanting to see head nods more than other audience behaviors. In addition, presenters were less interested in seeing negative states such as sleepiness or sadness as compared to more positive states such as engagement or excitement, possibly because negative feedback may indicate lack of audience interest that might induce performance anxiety associated with giving formal presentations [10]. It is important to note, however, that these findings could vary depending on the type of the presentation (e.g., teaching, sales) which was not captured by our survey. Figure 2 illustrates the full list of responses and their mean ratings.

Finally, participants felt the most comfortable sharing their video feed in smaller audience sizes. In particular, 84.5% of the participants mentioned that they would consider sharing their video in audience sizes with less than 5 people, 79.4% in audiences between 5 and 10, 64.5% in audiences between 11 and 20, 51.5% in audiences between 21 and 50, 45.3% in audiences with more than 50 people. Participants also indicated that some of the most frequent factors that would prevent them from activating their camera would be internet bandwidth issues, distractions in their background, and other people not having their videos turned on.

## 4 SYSTEM

Our findings from the exploratory survey indicated the requirement to see feedback from the audience as well as potentially useful non-verbal responses. This section describes the system design in terms of sensing, spotlighting behavior, and integration with a current videoconferencing application to act as a test bed for our evaluation study.

### 4.1 Sensing

As a proxy to capture the largest number of cognitive states and behaviors presented during the exploratory survey, we leveraged state-of-the-art affect-sensing computer vision algorithms. Given the video feed of an audience member, we extracted the following types of information:

**Face and Landmarks.** We used the Microsoft Face API[3] to detect the faces in each of the video frames and applied a landmark detector to identify relevant face areas (e.g., eyes, nose, mouth) as well as head pose orientation (e.g., yaw, roll), which were then used to extract additional types of information.

**Facial Expressions.** If a face was detected in a given video frame, the image region, defined by the Face API bounding box, was cropped and input into a Convolutional Neural Network (CNN) facial action unit classifier which was used to estimate the facial expressions. More details of the expression detection algorithm can be found in [38] which were externally validated in [35]. From the available set of expression categories, namely anger, disgust, fear, happiness, sadness, surprise, and neutral, we selected the ones that more closely matched the states presented during the exploratory survey such as sadness, neutral, happiness and surprise. As confusion was not amongst the available states and was highly ranked in the survey, we developed an additional neural network classifier that detected brow furrowing expression (Action Unit 4 in the Facial Action Coding System (FACS) taxonomy [39]) which is commonly shown during confusion. The same model architecture as in the expression classification described above was used in the process. The classifier was validated on the DISFA dataset [40] yielding an F-1 score of 70.2%. For each of the metrics, the models provided a probabilistic confidence value indicating the absence (0) or presence (1) of a certain expression.

**Head Gestures.** We used a Hidden Markov Model (HMM) [41] to calculate the probabilities of the head nod and head shake gestures. In particular, the HMM used the head yaw rotation value to detect head shakes, and the head Y-position of the facial landmarks to detect head nods over time. More details about the algorithm and validation can be found in [35].

Figure 2 highlights the cognitive and behavioral information (orange bars) that were directly or partly captured by some of the metrics above. Future work will consider the development of additional components that more comprehensively capture audience responses. The computer vision models were implemented on top of the *Platform for Situated Intelligence*[4] which facilitates the analysis of video data in real-time [42].

### 4.2 Spotlighting Behavior

To identify the most reactive face to display, we computed a score for each video frame of every audience member. In particular, we computed a weighted average of the sensed metrics for each frame, in which the weights were adjusted to closely resemble the responses of the exploratory survey findings and further refined during a pilot evaluation. In our case, less preferred responses indicated in the survey such as sadness and neutral received lower weights (below 0.1), and more preferred responses such as confusion and head-nods received higher weights (above 0.5). The score of each audience member was accumulated over non-overlapping periods of 15-seconds. At the end of each period, the audience member with the highest cumulative score was spotlighted for the next 15 seconds. If several members had the same score, the system would randomly select one of the highest, with the constraint that the same member cannot be spotlighted twice in a row, to resemble the feeling of 'reading the room'. This cycle was then repeated until the end of the presentation. The selection of 15 seconds was informed by prior interface design guidelines that similarly explored the display of human faces [43]. In addition, we validated the time window during a pilot evaluation to ensure that the system could spotlight as many relevant behaviors as possible, while avoiding being too distracting. As this system was mostly designed to be presenter facing, we ensured that the presenter was never spotlighted. Figure 3 shows an overview of the main processing steps. The pseudocode for Spotlight Behavior is described in Algorithm 1.

### 4.3 Microsoft Teams Integration

To provide an experience as close to a real-world application as possible, we created a Microsoft Teams Bot based on a publicly available sample on GitHub[5]. The bot works by acting as an additional participant in the meeting. Therefore, the bot can "see" all of the incoming video from the other participants as well as send its own video to the meeting attendees. While the bot's outgoing video stream shows up alongside all of the other participants' video streams by default, we requested the presenter to "pin" the bot's video stream so it would occupy the entire Teams window. In that way, the presenter would only see the bot's outgoing video as well as the slides. The bot was implemented in C# as a .NET Azure Cloud Service running on an Azure Virtual Machine.

---
[3] https://azure.microsoft.com/en-us/services/cognitive-services/face/

[4] https://github.com/Microsoft/psi

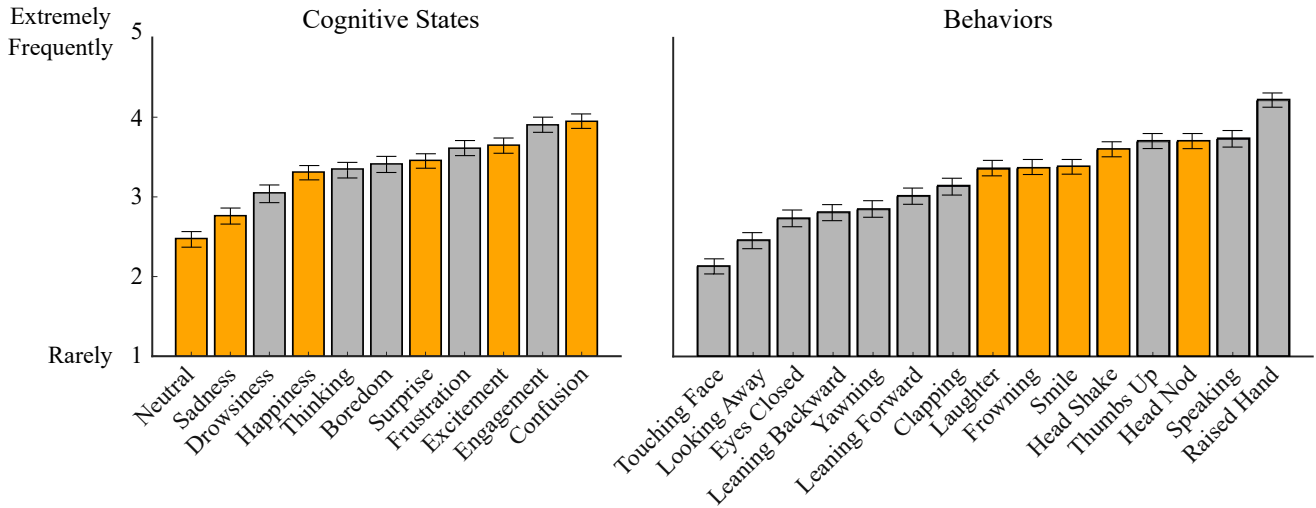

Figure 2: Presenters' preferences of audience reactions and cognitive states during online presentations. The error bars reflect the standard error. Orange bars indicate signals that were partially captured by AffectiveSpotlight.

---

**Algorithm 1:** Pseudocode for the Spotlighting Behavior

**Input:** Set of video feeds of audience members.
Initialize CurrentSpotlight to none
**while** Presentation **do**
  Initialize Timer
  Initialize Score to zero for each audience member
  **while** Timer is less than 15 seconds **do**
    **for** a in Audience **do**
      $Score_a \mathrel{+}= \sum_i^{Metrics} Weight_i * Get(Metric_i, a)$
  NextSpotlight ← member with highest Score. Break ties randomly **if** NextSpotlight is CurrentSpotlight **then**
    NextSpotlight ← member with second highest Score
  CurrentSpotlight ← NextSpotlight
  Bot displays CurrentSpotlight

---

## 5 EVALUATION STUDY

To understand the effectiveness of AffectiveSpotlight in communicating audience reactions and making presenters aware of the audience, we conducted a controlled within-subjects experimental study, where we compared the use of AffectiveSpotlight to the use of two other baseline systems as control conditions. This section describes the experimental details as well as the studied population.

### 5.1 Protocol

Participants with diverse presentation experience level (more details in the next section) were invited to join a videoconference call on the Microsoft Teams platform, in which one of the participants was randomly selected to be a presenter and the others were selected to be audience members. All participants were instructed to follow camera guidelines to reduce potential errors introduced by non-ideal conditions for the computer vision algorithms. In particular, they were asked to set a neutral background on their call, ensure that their face was around the middle of the frame, have minimum face occlusions (no hats, sun glasses), have appropriate lighting conditions, as well as ensure their camera captured a frontal view of their faces, as shown in Figure 4.

After the initial setup, presenters were instructed to prepare and give three presentations on three pre-selected topics related to the COVID-19 pandemic. The topics were chosen to be of general interest. In particular, they were: "Should we wear masks at this time or not?," "Should we adopt outdoor dining at this time or not?," and "Should we play sports at this time or not?." To facilitate the preparation of the presentations, we provided a template of three slides containing the pros, cons, and a potential personal verdict of the specific topic that presenters had to modify. Presenters were given around 8 minutes before each talk to prepare and were encouraged to present for around 5 minutes. To evaluate the level of attentiveness by the audience members during the talk, the audience members were asked to answer questions about the content from the presentation after each talk. In addition, the audience members were instructed to mute their microphones during the presentation and to not interrupt the presenter via other means such as chat messages.

---
[5]https://github.com/microsoftgraph/microsoft-graph-comms-samples/tree/master/Samples/V1.0Samples/LocalMediaSamples/AudioVideoPlaybackBot

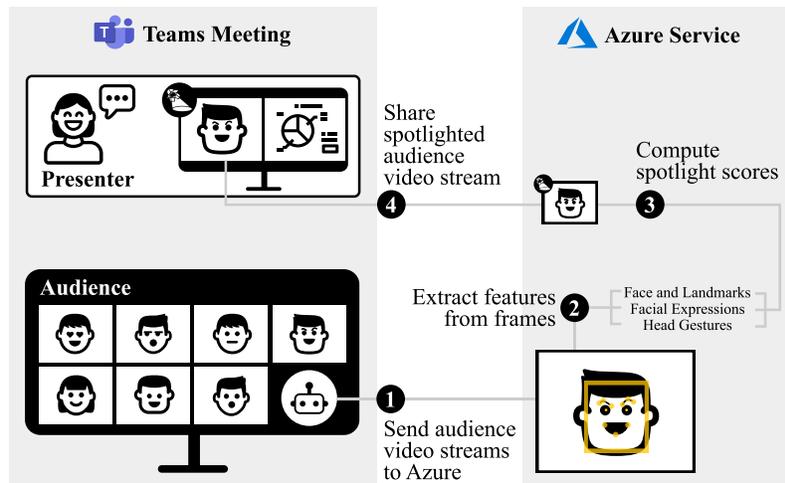

Figure 3: Overview of the system architecture and main processing steps: 1) the video of audience members are captured with a Microsoft Teams bot, 2) video feeds are analyzed to extract affective information in real-time, 3) scores for each audience members are computed and accumulated over a 15-second window, 4) the audience with the highest score is shown as the video feed of the bot.

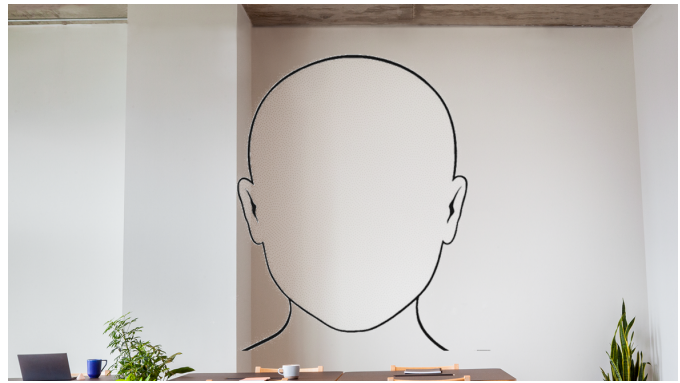

Figure 4: Background and camera orientation guidelines to improve AI performance

Presenters delivered each talk by using a different feedback support system in counterbalanced order to minimize ordering effects. In particular, the three feedback support systems were:

- **AffectiveSpotlight**: When in this condition, the presenters would see the slides on the right side of the screen, and the affectively-selected audience member on the left side (see Figure 1 right). Note, however, that the presenters were blind to the spotlight selection criteria.
- **RandomSpotlight**: When in this condition, the presenters would see the slides on the right side of the screen, and the randomly-selected audience member on the left side (same as Figure 1 right). Similar to the above case, presenters in this condition were blind to the spotlight selection criteria.
- **DefaultUI**: When in this condition, the presenters would see the default Microsoft Teams UI that predominantly shows the slides and a limited set of audience members at the bottom of the screen (see Figure 1 left). This condition allowed us to compare against the default user interface of an existing videoconferencing software.

The assignment of the topics was randomized to minimize potential topic effects. Also, none of the audience members were informed about the details of the system capabilities and the conditions to keep their responses as consistent as possible. Finally, participants were asked to fill out several surveys throughout the study, which are described in detail in the next section.

This study protocol was reviewed and approved by the institution's ethics review board and is illustrated in the Figure 5. Each session lasted approximately 60 minutes, and participants of the study were compensated with a $30 or $25 gift card depending on whether they were presenters or audience members, respectively. All presentations were recorded for analysis as well.

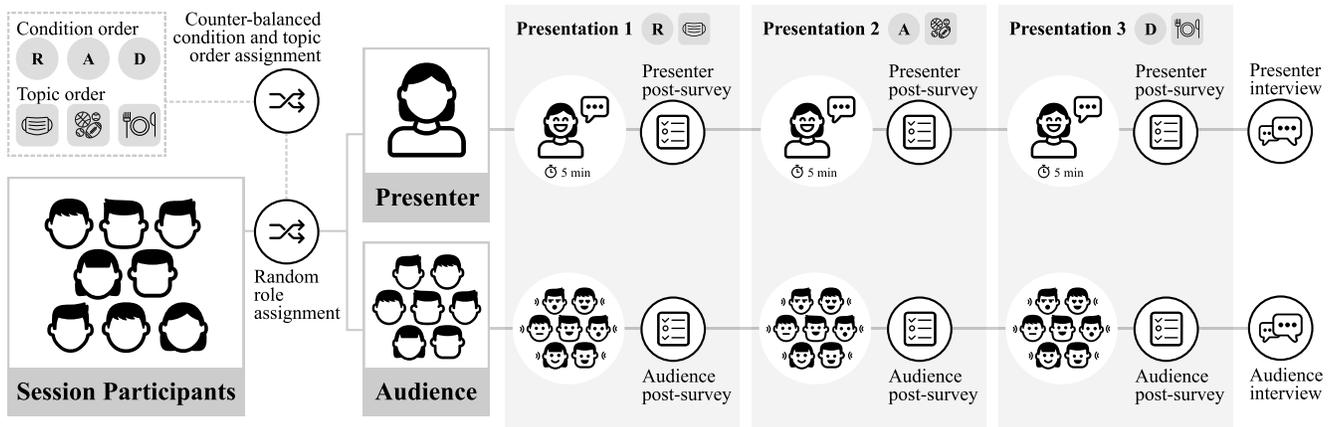

Figure 5: Overview of the experimental protocol. The presenter gave three talks with different support systems: AffectiveSpotlight (A), RandomSpotlight (R), and DefaultUI (D).

## 5.2 Methods

At the beginning of the study, both the presenter and audience members were asked to complete a survey for demographics information (e.g., age, gender) as well as the Self-Perceived Communication Competence Scale (SPCCS) [44] to capture self-reported competence over a variety of communication contexts.

After each of the three talks, presenters were asked to complete (1) a system evaluation survey [45] to capture the presenters' assessment of the feedback system as shown in Table 1, (2) an audience awareness survey [46] to capture the presenter's experience with the tool in gauging the audience responsiveness and reactions as shown in Table 2, and (3) an assessment of the quality of the presentation [47] to capture the presenter's perception of their talk as shown in Table 3. Similarly, audience members were asked to complete (1) an assessment of the quality of the presentation [47] to capture the audience's perception of the talk as shown in Table 4, and (2) open-ended questions related to the content of the presentation. In particular, we asked "Share one pro mentioned by the speaker," "Share one con mentioned by the speaker," and "What was the personal verdict of the speaker about the presentation topic?" Finally, both presenters and audience members participated in a semi-structured interview to provide qualitative feedback about the system at the end of the study.

To analyze the data, we followed a mixed-methods approach. We analyzed the self-reports and videos to obtain data for quantitative analysis, and the lead author of this submission open-coded our interviews with the presenter (P) and the audience (A) for qualitative analysis [36, 37]. To assess the potential significant differences on survey responses across conditions, we performed Paired Wilcoxon signed-rank tests with Bonferroni corrections for all survey questions, unless mentioned otherwise.

## 5.3 Participants

We recruited a total of 117 participants. From these, 14 participants were selected to be presenters, with 8 male and 6 female participants and a mean age of 37 years old (min=21, max=60). Presenters were pre-screened to be equally split into having high or moderate public speaking competence, according to the SPCCS measure [44]. The rest of the participants were assigned the audience role, with 52% males (48% females) and a mean age of 33 years old (min=21, max=60). Informed by the findings from our exploratory survey where we identified typical online meeting sizes to be between 5 - 10 people and the expression of high comfort levels in enabling the web camera in meetings of the same size, we set the average group size to be 8 people, including the presenter.

## 6 RESULTS

This section provides results focused on the behavior of the system, followed by its evaluation and then continues by analyzing the impact of the different support systems in terms of audience awareness, quality of the presentation, and presenters' anxiety.

### 6.1 How was the system perceived by the presenters?

Table 1 shows the average responses for the questions focused on the evaluation of the platform by the presenters. The *AffectiveSpotlight* was rated significantly higher than the other two conditions (*RandomSpolight* and *DefaultUI*) in terms of ease-of-use ($Z = -2.40, p = 0.008; Z = -2.70, p = 0.006$), platform satisfaction ($Z = -2.74, p = 0.008; Z = -2.68, p = 0.012$), and likely future use ($Z = -2.31, p = 0.031; Z = -2.98, p = 0.011$). In terms of how much the platform helped deliver the presentation, the *AffectiveSpotlight* was also rated significantly higher than the *RandomSpotlight* ($Z = -1.80, p = 0.005$). No significant differences were found across the different conditions in terms of anxiety and distraction elicited by the platform.

These findings seem consistent with the qualitative feedback provided during the interviews. For instance, one participant stated *"the [AffectiveSpotlight] system was useful when presenting"* [P1], another one said the *"[AffectiveSpotlight] made it less challenging for them to see audience responses"* [P6], and *"[AffectiveSpotlight] increased the overall feeling of presenting"* [P11]. In terms of perceived performance of the system, several presenters mentioned that the spotlight interface was *"intuitive to process"* [P1] and they could *"process and see reactions from the audience as they were speaking"* [P10]. The use of a face, rather than attaching emotional labels to the audience, made them feel *"closer to presenting in person"* [P9], *"helped them pick subtle non-verbal cues more easily"* [P14] and *"made it more like a physical presentation"* [P12]. Some presenters also expressed concerns that *"the system could bias towards certain facial constructs and personality types"* [P11] and *"end up focusing on one or two people, which is not helpful"* [P12]. They indicated that an injection of *"randomness"* [P2], along with spotlighting *"the more reactive members"* [P5] could be more helpful for them than spotlighting only the reactive audience members.

| Question with endpoints: "Not at all" (1) and "Very Much" (7) | AffectiveSpotlight | RandomSpotlight | DefaultUI |
|---|---|---|---|
| How easy to use was the platform? | **6.40 (0.74)** | 5.50 (1.34) | 5.50 (0.94) |
| How anxious did you feel when using the platform? | 3.67 (2.06) | **3.79 (1.72)** | 3.07 (1.21) |
| How much do you feel the platform helped you deliver the presentation? | **5.33 (1.50)** | 4.14 (1.23) | 4.93 (1.69) |
| How distracting was the platform when delivering the presentation? | **4.13 (1.64)** | 3.86 (2.18) | 3.86 (1.51) |
| How satisfied are you with the platform? | **5.73 (1.10)** | 4.93 (1.21) | 4.14 (1.29) |
| How much would you like to give future presentations with the platform? | **5.87 (1.36)** | 4.71 (2.02) | 5.00 (1.52) |

**Table 1: Average and standard deviation for the system evaluation survey.**

## 6.2 What was the impact on presenters' awareness of the audience?

The analysis of the video recordings showed that there were around 20 spotlight changes for each of the *RandomSpotlight* and *AffectiveSpotlight* conditions. While 87% of the audience members were showed in the random condition, only 40% of them were shown during the *AffectiveSpotlight*, highlighting that affective responses were mostly displayed by a subset of the audience members. In the following, we analyze the average responses for the questions focused on the presenter's awareness of the audience members (see Table 2).

The *AffectiveSpotlight* received significantly more positive responses than the other two conditions (*RandomSpolight* and *DefaultUI*) in terms of awareness of presentation performance ($Z = -2.37$, $p = 0.017$; $Z = -2.13$, $p = 0.033$), ease-to-see audience responses ($Z = -2.17$, $p = 0.028$; $Z = -2.23$, $p = 0.026$), and ease-to-respond to audience responses ($Z = -2.46$, $p = 0.014$; $Z = -2.19$, $p = 0.028$). In addition, both the *AffectiveSpotlight* and *DefaultUI* received significantly more positive responses than the *RandomSpotlight* in terms of personal connection with the audience ($Z = -2.35$, $p = 0.019$; $Z = -2.44$, $p = 0.015$).

These findings seem to be further supported by qualitative feedback in which presenters described achieving more awareness of the audience with *AffectiveSpotlight*. For instance, one participant stated *"whatever energy I'm putting in, it seems like it's getting reflected"* [P4]. Another participant mentioned that the AffectiveSpotlight helped *"create a feedback loop with the audience"* [P2] that enabled them to understand the audience better: *"So you know biofeedback of sorts, so it's like, oh OK, if I do this then these people smile and those people laugh or whatever"* [P10]. Regarding the mapping between different audience facial expressions and the instantaneous presenter reactions, we observed that the 'head-nodding' of audience members made one presenter feel *"validated"* [P3], the facial displays of confusion *"threw off [one presenter]"* [P6] but also "encouraged [another presenter] adapt the talk" [P8], and the displays of audience smiling/laughing helped another presenter determine that *"the points were landing well"* [P9].

In addition to these, presenters also reported the need for online systems that can *"capture and create that personal connection between a presenter and participants that is achieved in physical talks"* [P8] and felt that the AffectiveSpotlight system helped achieve a *"connection with the audience"* [P7] or enabled them to *"be better in tune with the others [audience]"* [P9] in the meeting. Finally, some presenters also indicated that they wished to know *"how the spotlight works"* [P8] and *"how it picks people"* [P3] to *"actually get a better picture of all the audience members"* [P12] and *"become aware of their performance"* [P3].

## 6.3 What was the impact on the quality of the presentation?

Table 3 and Table 4 show the average rating of the quality of the presentation for presenters and audience members, respectively. We did not find any significant differences in these ratings between the conditions, suggesting that the perceived quality of the talks was consistent for all the conditions.

Motivated by the findings of the previous section, we wanted to further explore whether there were some similarities between the presenter and audience ratings. To do so, we created a similarity score for each of the three sessions for every presenter, which was computed as the absolute difference between the presenter rating and the average of the audience ratings for each of the sessions. Therefore, a lower similarity score indicates that the presenter self-reports are more closely aligned with those provided by the audience members. Considering this score, responses in the *AffectiveSpotlight* were found to be significantly more similar than the other two conditions (*RandomSpolight* and *DefaultUI*) in terms of both satisfaction with the presentation ($Z = -2.097$, $p = 0.036$; $Z = -2.13$, $p = 0.022$) and overall engagement ($Z = -2.86$, $p = 0.004$; $Z = -2.53$, $p = 0.02$).

These findings seem consistent with qualitative reports that emphasized that recreating the notion of "feel the room" helped presenters understand the audience members, especially during the high-demanding cognitive task of public speaking. One participant mentioned *"I just somehow in the second one [AffectiveSpotlight],*

| Question with endpoints: "Not at all" (1) and "Very Much" (7) | AffectiveSpotlight | RandomSpotlight | DefaultUI |
|---|---|---|---|
| How much of a personal connection did you feel with the audience? | **5.20 (1.37)** | 3.93 (1.64) | 4.93 (1.38) |
| How aware were you of your presentation performance? | **5.47 (1.30)** | 4.29 (1.14) | 4.43 (0.94) |
| How easy was to see the non-verbal feedback from the audience? | **5.87 (0.83)** | 4.71 (1.82) | 5.00 (1.36) |
| How easy was to respond to the non-verbal feedback from the audience? | **5.53 (1.13)** | 4.36 (1.69) | 4.43 (1.34) |

Table 2: Average and standard deviation for the audience awareness survey.

*somehow I felt that people were giving me like a good amount of time to sort of focus with them. I realized that OK, I'm now engaging or not, and sort of you adapt"* [P15]. Although the presenters mentioned that the system provided them with certain cues that helped them make changes to their talk, they also expressed concerns that *"they might potentially get thrown off if the reaction from audience violates their expectation"* [P6], or *"if they did not know how to react"* [P7]. In general, audience members also felt that the presenters were cognizant of the reactions provided to them and they adapted their talks to it: *"when I was not convinced about the cons [of a particular slide] in the second talk [AffectiveSpotlight], I felt the presenter spent a lot of time trying to get their point unlike the first talk [RandomSpotlight]"* [A51] and *"the presenter tried to emphasize more in one of the talks [AffectiveSpotlight]"* [A82], identifying that the audience felt changes in presenter's delivery of the talk when using the AffectiveSpotlight. On the contrary, one audience member mentioned that *"the presenter kept joking because one or two people found it funny and that was slightly off putting"* [A23] and that in a regular meeting they *"would have liked to tell them to keep it moving by just speaking up but they could not really indicate that anyhow here"* [A23].

### 6.4 What was the impact on the presenter's anxiety?

Although no significant differences were found when reporting the potential anxiety elicited by the platform (more details in Section 6.1), a recurring theme across the qualitative feedback was focused on the relationship between audience responsiveness and presenter's anxiety. In particular, several presenters mentioned that the AffectiveSpotlight helped with their *"anxiety because at various points during the talk you are not really sure how you are resonating [with the audience] and you are like always guessing"* [P9], and *"it helped not constantly think about it [audience reactions to the talk] at the back of my mind"* [P11]. The availability of technology to *"do the job of panning the room, so as to speak"* [P14] made them feel *"not having to withdraw when audience is disengaged"* [P7] but rather *"continue to communicate with those people"* [P9]. Finally, AffectiveSpotlight seemed to encourage presenters to *"put some extra information beyond the slides"* [P2], and one presenter mentioned that *"I tried to act a little more silly, to see if they actually laughed"* [P8].

These findings were further supported when we examined the talk duration across conditions. One-way within-subjects ANOVA test revealed a significant differences across conditions in terms of talk duration ($F(2, 40) = 4.013; p = 0.026$). While presenters of the study were recommended to speak for the same amount of time for all the conditions, we observed that presenters in the *AffectiveSpotlight* spoke for more seconds ($M = 264.17, SD = 95.51$) than the *RandomSpotlight* ($M = 202.27, SD = 82.47$) and the *DefaultUI* ($M = 207.25, SD = 64.96$). These findings are also consistent with previous research showing that higher audience responsiveness is positively correlated with greater willingness to speak [10, 48–50].

## 7 DESIGN RECOMMENDATIONS

Our findings revealed that appropriately spotlighting the audience responses during an online presentation can improve the presenter's presentation experience, encouraging them being aware of the audience which is crucial to "reading the room." Based on these findings, the design choices that our participants resonated with, and qualitative suggestions provided by our participants, we identified several key considerations for the design and development of real-time audience feedback systems.

### 7.1 Accounting for the limitations of the AI systems according to the context of use

During a pilot evaluation, we observed large variability in the use of video backgrounds, lighting conditions, and camera angles which negatively impacted the performance of the computer vision algorithms. Despite the recent progress in AI, it is important to note that sensing algorithms are still far from perfect. In our study, we employed several countermeasures to ensure high performance. We asked participants to follow a set of camera calibration guidelines (Section 5.1) to maximize the performance of the sensing algorithms during the analysis of their video feeds. In addition, we avoided the explicit labeling of expressions to minimize the potential impact of misclassifications. Instead, we used AI to influence the signals provided to the presenter so they could more effectively interpret the data based on prior experience and contextual information. This approach is particularly relevant in the context of emotion recognition in which the subjective experience and expression of emotion can vary significantly from person to person [51].

### 7.2 Allowing the presenters and the audience to control the behavior of the system

During the semi-structured interviews with the participants of our study, both presenters and audience members expressed a strong desire to have more control over the AffectiveSpotlight. On the one hand, presenters wanted the ability to select the sensed metrics to better achieve individual goals. For instance, one presenter stated

| Question with endpoints: "Not at all" (1) and "Very Much" (7) | AffectiveSpotlight | RandomSpotlight | DefaultUI |
|---|---|---|---|
| How satisfied are you with the presentation? | **5.78 (0.97)** | 4.64 (1.44) | 5.46 (1.59) |
| How engaging was the presentation? | 4.71 (1.85) | 3.92 (1.27) | **4.80 (1.42)** |
| How nervous were you during the presentation? | 3.00 (2.14) | **3.64 (2.13)** | 3.06 (1.83) |
| What is the overall quality of the presentation? | 4.83 (0.91) | **5.00 (1.30)** | 4.78 (1.31) |

Table 3: Average and standard deviation for the presenter's self-evaluation of the presentation.

| Question with endpoints: "Not at all" (1) and "Very Much" (7) | AffectiveSpotlight | RandomSpotlight | DefaultUI |
|---|---|---|---|
| How satisfied are you with the presentation? | 5.59 (1.38) | **5.81 (1.14)** | 5.64 (1.27) |
| How engaging was the presentation? | 5.09 (1.64) | **5.13 (1.40)** | 5.08 (1.49) |
| How nervous were you during the presentation? | 5.18 (1.47) | **5.20 (1.42)** | **5.20 (1.44)** |
| What is the overall quality of the presentation? | 4.18 (1.62) | 3.92 (1.55) | **4.38 (1.55)** |

Table 4: Average and standard deviation for the audience evaluation of the presentation.

*"seeing who is raising a hand or speaking during Q/A sessions"* [P3], and another one said that *"if I am explaining a purely complicated topic, it would be helpful to only focus on the confused audiences"* [P5]. Further, presenters mentioned that seeing positive and negative reactions from the audience could impact their presentation confidence and flow differently and would thus like to control the behavior of the spotlight accordingly. One presenter mentioned that *"seeing positive non-verbal reactions from the audience as opposed to furrowed brow gives them confidence"* [P1], while another mentioned that *"sometimes if people don't get what I am saying or [they are] acting confused, I get thrown off."* [P7]. On the other hand, audience members also expressed interest in influencing the sensed metrics and/or the possibility to provide explicit feedback to change the behavior of the spotlight. Negotiating the level of control and agency between human-AI collaboration is well-known design challenge from prior research [52, 53]. Control of this kind could also benefit speakers how might want to moderate sensory inputs in order to reduce the stress of video-mediated interactions [54].

### 7.3 Being transparent about the system's capability and use

To help facilitate consistent audience responses throughout the study, participants were not fully pre-informed about the specifics of each experimental condition. However, when considering real-life deployments and adoption, it is important to communicate when and for what purpose the system may be used. Similar to existing online videoconferencing capabilities such as recording, presenting or sharing screen, we envision the proposed system would need to be triggered by the presenter and accepted by the audience to help provide feedback during the presentation. In addition, it is important to let the audience know when and what information is being provided to the presenter. To further promote transparency, future designs could explore providing personalized notifications to each spotlighted member and evaluate whether such designs would make the audience members more self-conscious or display less natural behavior.

### 7.4 Respecting the privacy of the audience members

Even though facial expressions and head gestures may not necessarily represent the internal state of people [55], many people still consider these types of information to be personal and private. Consequently, the deployment of such systems may trigger polarized views in users. Despite the positive feedback in our study, a small percentage of participants mentioned that they would rather not contribute to the audience feedback process as they may be multi-tasking and/or find video analysis to be invasive. Our exploratory study indicated that this hesitation may be less pronounced when considering smaller audiences, but we envision different people may prefer different levels or types of engagement. For instance, this work only considered implicit feedback methods to sense the audience response, but more explicit feedback methods (e.g., self-reports) may be preferable for those who are more concerned about privacy. These could help recreate the "sitting in the back of the room" experience in which audience members may be more passively involved. Finally, several participants emphasized their preference to have emotion sensing algorithms to run locally so that their images were never affectively-analyzed on the cloud.

## 8 DISCUSSION

This work addresses the problem of analyzing and facilitating audience feedback during online presentations. Through an exploratory survey, we first captured presenters' preferences for the audience cognitive states and behaviors that they would like to see when presenting, such as confusion and head nods, which helped us design and develop our presenter support system. Inspired by efforts in cinematography and HCI, we created AffectiveSpotlight, which is a Microsoft Teams bot that analyzes audience's facial expressions and head gestures, and spotlights audience members in real-time to the presenter.

To evaluate the system, we performed a within-subjects study in which 14 presenters gave talks to groups of around 8 people (total of 117 participants) with the AffectiveSpotlight and two other control support systems: a randomly-selected spotlight and the default Microsoft Teams UI. When evaluating the system, we found that presenters provided significantly more positive ratings when using the AffectiveSpotlight than the other two systems in terms of system satisfaction, ease-of-use, and future potential use. In addition, presenters provided significantly higher ratings when experiencing the affectively-selected spotlight vs. the randomly-selected spotlight in terms of how much the platform helped them give the presentation, suggesting that the content of the spotlighted information was found to be helpful. More importantly, we found that presenters in the *AffectiveSpotlight* condition reported to be significantly more aware of the audience. Several of our presenters described it as achieving a communicative feedback-loop that enabled them to adapt their presentation as needed (e.g., injecting a joke, provide clarifications) based on the audience reactions. Similar methodology has previous been successfully evaluated in the context of emotion regulation [33, 56].

Further, the increase of audience awareness may have also influenced presenter's own evaluation of the quality of their talk, as indicated by a stronger similarity between the self-reported evaluation of their presentation quality with that of the audience members when using the AffectiveSpotlight. Finally, our qualitative analyses suggested that the AffectiveSpotlight had an effect on the presenter's anxiety and confidence, that is typically impacted by lack of audience feedback in online presentations. We found that presenters spoke for a significantly longer period of time when using the AffectiveSpotlight, potentially indicating a reduction in attempt to withdraw from the speaking situation, upon access to audience responsiveness [10, 50]. Overall, these findings seem to support that the proposed support system empowered online presenters to access audience reactions real-time, and make online presentations a bit closer to live presentations in the context of Motley's continuum [12]. To help facilitate the development of future systems in this space, we also identified four recurrent design recommendations based on the design, development, and evaluation of our system. In particular, we highlighted some of the challenges and potential solutions around current limitations of AI, user control, system transparency, and data privacy.

## 9 LIMITATIONS

Despite the positive results, it is important to note that there are several limitations in this study. Although the exploratory survey captured the preference of presenters to see a set of audience reactions and cognitive states, we did not explicitly ask how their preferences may change for various types of presentations (e.g., teaching, sales), which may have led to different findings. The design choice for the spotlight to show certain reactions and cognitive states is thus dependent on the specific use cases and presentation types, and our recommendation is that presenters have control over the design criteria.

Due to the exploratory nature of our work, the presenters and audience members had no previous relationships to reduce potential familiarity/preference biases. However, familiarity could determine the spotlight design choice as well as the preference for presenters to prioritize certain audience reactions more. For instance, a presenter may want to focus on familiar supportive faces to help alliviate stress and a teacher may want to focus on confused students to help provide timely support. These factors could have an impact on the spotlight design choice as well as the generalization of the findings across various presentation scenarios.

In addition, although we recruited a total of 117 participants in our evaluation study, they were grouped into 14 groups which reduced the sample size of studied presenters. Furthermore, all the participants were recruited from the same technology company which represents a biased and limited set of the population. Moreover, to help start exploring the potential utility of spotlights, this study considered a controlled environment in which presenters were asked to give short and curated talks which may offer limited ecological validity when considering real-life presentations. To address this, future usability work will need to consider real-life speaking engagements with different topics that more naturally trigger different levels of public speaking anxiety, as well as sustained used of the feedback system to quantify potential novelty effects. The potential deployment of a spotlight technology should also account for challenges highlighted by the need for transparency (e.g., use of consent) and privacy (e.g., AI processing taking place locally) as indicated in section 7. Despite these limitations, however, we found several significant findings which highlight the potential value of research in this space.

## 10 FUTURE WORK

This work has also helped identify relevant opportunities for future research. The proposed system considers the measurement of head gestures and facial expressions to identify the most reactive audience members during a specific time window. In our work, we empirically set the refresh window to be 15 seconds which helped us avoid potential distractions, but could have also missed relevant responses. To address this, we envision future spotlight systems may consider a flexible window that can more quickly reflect the behaviors as they occur. In addition, some of the participants of the study expressed strong interest in controlling the specific behavior of the spotlight which could have helped address specific needs (e.g., detecting confusing points). As we consider expanding the sensing modalities (e.g., microphones, eye tracking), we believe that the quality of the information provided by the spotlight can be improved. However, it is important to be mindful that different sensing channels may be considered more invasive than others. We recognize the comfort levels expressed in sharing video feeds more in smaller meetings and thus propose several directions for future system use. For instance, the audience could be sensed locally, and their video be provided only to the presenter. In addition, the proposed system also extends to using affect sensing for sampling strategies in existing video conferencing systems, where only a subset of audience is shown to the presenter. Our interviews also show promise in combining both explicit and implicit feedback methods to more effectively address the preferences of different audience members. Finally, we believe future work will need to focus on the development of new user interfaces that promote system transparency in terms of both capturing and reflecting audience feedback which will be critical to prevent potential misuses and maximize user adoption.

## 11 CONCLUSIONS

This work introduces AffectiveSpotlight, a real-time feedback support system for online presenters that analyzes and spotlights audience members based on their affective responses. Informed by an exploratory survey and evaluated in a controlled within-subjects study, we demonstrated some of the potential benefits of facilitating non-verbal audience feedback via the proposed system versus two other control support systems. We hope our findings and design recommendations will help enable future work exploring the possibilities of affect sensing and AI-mediated interactions in the context of online meetings. We are looking forward towards a future when similar approaches can continue to enhance presentation experiences and help close the gap between online and in-person presentations.

## ACKNOWLEDGMENTS

The authors would like to thank Robert Sim, Piali Choudhury, and Shane Williams for their help in the development of scripts for the user study and the bot. We would also like to thank Jaime Teevan for her initial feedback on the concept. We thank Samiha Samrose for providing us feedback at various stages of the project and evaluating the initial prototypes. Finally, we thank the participants of our studies for their time in evaluating the concept and the prototype.